\documentclass[10pt,a4paper,twocolumn]{article}
\usepackage{f1000_styles}

\usepackage[numbers]{natbib}
\usepackage{url}

\usepackage{authblk}
\usepackage{footmisc}
\usepackage{url}

\begin{document}

\title{Design choices for productive, secure, data-intensive research at scale in the cloud.}
\author[1]{Diego Arenas }
\author[2]{Jon Atkins}
\author[9]{Claire Austin}
\author[2]{David Beavan }
\author[2, 8]{Alvaro Cabrejas Egea}
\author[10]{Steven Carlysle-Davies}
\author[2]{Ian Carter}
\author[4]{Rob Clarke}
\author[6]{James Cunningham}
\author[ ]{Tom Doel}
\author[2]{Oliver Forrest}
\author[2]{Evelina Gabasova}
\author[2]{James Geddes}
\author[2]{James Hetherington \thanks{Corresponding author: jhetherington@turing.ac.uk}}
\author[2]{Radka Jersakova }
\author[2]{Franz Kiraly }
\author[2]{Catherine Lawrence}
\author[2]{Jules Manser}
\author[2]{Martin T. O'Reilly }
\author[2]{James Robinson }
\author[13]{Helen Sherwood-Taylor}
\author[5]{Serena Tierney}
\author[2, 3]{Catalina A. Vallejos }
\author[2, 7]{Sebastian Vollmer}
\author[2, 4]{Kirstie Whitaker }
\affil[1]{ 
The University of St Andrews, KY16 9SX, Scotland.}
\affil[2]{The Alan Turing Institute, 96 Euston Rd, Kings Cross, London, NW1 2DB, UK.}
\affil[3]{MRC Human Genetics Unit, Institute of Genetics \& Molecular Medicine, University of Edinburgh, Western General Hospital, Crewe Road, Edinburgh, EH4 2XU.}
\affil[4]{Department of Psychiatry, University of Cambridge, Herchel Smith Buidling for Brain \& Mind Sciences, Forvie Site, Robinson Way, Cambridge CB2 0SZ, UK.}
\affil[5]{VWV LLP, 24 King William Street, London EC4R 9AT.}
\affil[6]{Division of Informatics, Imaging and Data Sciences, School of Health Sciences, Faculty of Biology, Medicine and Health, The University of Manchester, Oxford Road, Manchester, M13 9PL}
\affil[7]{Warwick Mathematics Institute \& Department of Statistics, University of Warwick, CV4 7AL}
\affil[8]{MathSys Centre for Doctoral Training, Warwick Mathematics Institute, University of Warwick, CV4 7AL}
\affil[9]{The British Library, 96 Euston Road, London NW1 2DB}
\affil[10]{Edinburgh Parallel Computing Centre, University of Edinburgh, Bayes Centre, 47 Potterrow, Edinburgh, EH8 9BT}

\maketitle
\thispagestyle{fancy}


\begin{abstract}

We present a policy and process framework for secure environments for productive data science research projects at scale, by combining prevailing data security threat and risk profiles into five sensitivity tiers, and, at each tier, specifying recommended policies for data classification, data ingress, software ingress, data egress, user access, user device control, and analysis environments.
By presenting design patterns for security choices for each tier, and using software defined infrastructure so that a different, independent, secure research environment can be instantiated for each project appropriate to its classification, we hope to maximise researcher productivity and minimise risk, allowing research organisations to operate with confidence.

\end{abstract}

\clearpage

\section{Introduction}

\subsection{Scope}

Secure environments for analysis of sensitive datasets are essential for research.

Such ``data safe havens'' are a vital part of the research infrastructure.
It is essential that sensitive or confidential datasets are kept secure, both to comply with legal and contractual requirements and to retain the confidence and consent of data providers to continue to make their data available for use in research. Of particular importance is the capability to make use of personal data in a manner that is compliant with data protection law, 
and to avoid harm to the consent of society for research activities with personal data (called `social license').

To create and operate these environments safely and efficiently, whilst ensuring usability, requires, as with many socio-technical systems, a complex stack of interacting 
business process and design choices. Well-understood standards for organisations to follow 
exist at many levels of this stack, but we believe an important piece is missing. This paper therefore seeks to help the research e-Infrastructure community reach a new consensus on the further specific choices that need to be made when building and managing environments for productive, secure, collaborative research projects.

We propose choices for the security controls that should be applied in the areas of:

\begin{itemize}
\item data classification
\item data ingress
\item data egress
\item software ingress
\item user access
\item user device management
\item analysis environments
\end{itemize}

We do this for each of a small set of security ``Tiers'' - noting that the choice of security controls depends on the sensitivity of the data.

We acknowledge we are excluding a number of important areas that do need to be considered, so as to create a manageable body of work.

We do not cover the fundamental organisational security practices 
necessary for any organisation outside research (such as the NCSC's Cyberessentials Plus~\cite{Cyberessentials}). Nor do we cover the data-centre level or organisational management security practices which are fundamental to any secure computing facility. These are each essential areas of guidance. We will \emph{assume} these good practices are followed, and that an organisation already has 
access to a secure ISO 27001~\cite{ISO27001} compliant data centre, robust governance processes, and fundamental security practices for business technology.

A \emph{data facility} provides
persistent access to one or more secure datasets, and the compute needed to analyse them, over multiple projects and long periods of time. This challenge,
faced by libraries, cohort studies, and experimental facilities, is an essential one also to address, but it is not the challenge addressed here.

Lastly, we do not in this paper cover the ``how'': how do we actually build such an environment, constructing
it from software configuration files, firewall rules and so on? Our current reference implementation based on
Microsoft Azure will be the subject of a future paper.

\subsection{Approach}

\subsubsection{Secure data science}

Secure, productive environments for research must be co-designed by the research community and e-Infrastructure professionals.  A clear understanding of the requirements will help to avoid adversarial approaches, 
between those who feel greatest the pain of lost research productivity, and those who bear the risk of breach. 
This paper seeks to encourage a collaborative approach, building a common understanding of the framework within which usability-security trade-offs are made. 

We highlight three assumptions about the research user community critical to our design:

Firstly, we must consider not only accidental breach and deliberate attack, but also the possibility of `workaround breach', where
well-intentioned researchers, in an apparent attempt to make their scholarly processes easier, circumvent security measures, for example by copying out datasets to their personal device. This may happen where users regard such security measures as barriers to overcome rather than necessary processes, for example due to institutional "security fatigue" resulting from unnecessarily restrictive or ineffectual security measures. 
Our user community is relatively technically able, so the casual use of technical circumvention measures, not by adversaries but by
colleagues, must be considered.
This can be mitigated by increasing awareness and placing inconvenience barriers in the way of undesired behaviours, even if those barriers are in principle not too hard to circumvent, and by educational interventions, such as through requiring users to read appropriate data handling training materials as part of the access process. Security measures should be appropriate to the sensitivity of the data, so that users understand the purpose of the measures and how they relate to their personal responsibilities.

Secondly, research institutions need to be open about the research we carry out, and hence, the datasets we hold. This is because of both the need to 
publish our research as part of our impact cases to funders, and the need to maintain the trust of society, which provides our social licence. This means
we cannot rely on ``security through obscurity'': we must make our security decisions assuming that adversaries know what we have, what we are doing with it, and
how we secure it. 

Thirdly, academic users have a high degree of cultural autonomy. Institutional policies are never
perfectly followed in the highly distributed and open cultures necessary for research creativity. We believe we are morally responsible for breaches occurring as a result of researchers choosing not to use our facilities, despite organisational policies requiring them to do so. 
Thus, in providing a highly usable facility, nudging researchers to choose to use a shared secure environment, we reduce the opportunity for harm.

\subsubsection{Why Classify?}

One of the major drivers for usability or security problems is over- or under-classification, that is, treating data as more or less sensitive than it deserves.

Regulatory and commercial compliance requirements place constraints on the use of datasets. Implementation of that compliance must be set in the context of the threat and risk profile and balanced with researcher productivity.

Almost all security measures can be circumvented, security can almost always be improved by adding additional barriers, and improvements
to security almost always carry a cost in usability and performance.

Miss-classification is seriously costly for research organisations: over-classification results not just in lost researcher productivity, but also a loss of scientific engagement, as researchers choose not to take part in a project with cumbersome security requirements. Systematic over-classification \emph{increases} data risk by encouraging workaround breach.

The risks of under-classification include not only
legal and financial sanction, but the loss of the social licence to operate to the whole community of data science researchers\cite{carter2015social}.

\subsubsection{Software-defined infrastructure}

Our approach - separately instantiating an isolated environment for each project - is made feasible by the advent of ``software defined infrastructure''.

It is now possible to specify a whole arrangement of IT infrastructure, servers, storage, access policies and so on,
completely as \emph{code}. This code is executed against web services provided by infrastructure providers (the APIs
of cloud providers such as Microsoft, Amazon or Google, or an in-house ``private cloud'' using a technology such
as OpenStack~\cite{openstack}), and the infrastructure instantiated.

This paper \emph{assumes} the availability of a software-defined infrastructure provision offering, in an ISO27001-compliant data-centre and organisation, supporting the scripted instantiation of virtual machines, storage,
and secure virtual networks. 

One of the key pieces of functionality needed to support a data safe haven is ``Identification, Authorisation and Authentication'' (IAA). This is a complex part of the system that is critical to get right and we recommend relying on existing proven implementations. Many software-defined infrastructure offerings provide IAA as a service and we would suggest using this where available.

Our view is that a software-defined infrastructure platform, on which to build,
is a requirement for a well-defined secure research environment. It allows for separation of concerns between
core IT services and the research e-Infrastructure community.
It also means that the definition of the environment can be meaningfully audited - 
with as many aspects as possible of it described formally in code, it can be more fully scrutinised.

Given this, the topic of this paper is how best to assemble the pieces afforded by such services into 
a secure, scalable, productive research environment.

In this document, we do so at a high level, describing in general terms what should be built,
and in particular, what restrictions should be applied at which tiers.
In the followup reference implementation paper, we will give a detailed description
of a reference implementation on the Microsoft Azure public cloud, and publish the scripts
themselves.

Note that the use of a public cloud is not the critical aspect. The assumption is only
the presence of a programmable software defined infrastructure.
Whether the use of third-party data centre providers is admissible for
sensitive data is a matter of some debate within
the community, and our task in this paper is independent of that debate. Other work at the Turing Institute, notably the MARU project~\cite{maru}, addresses some of these concerns.

\subsubsection{Software defined management processes}

How do we make our \emph{business processes} software defined, so that the same code-level scrutiny
can be made of the processes that govern our activities as for the definition of the 
infrastructure?

Our recommendation is to
manage all the tasks that take place (user management, project creation, environment classification,
environment instantiation, data and software ingress and egress, and more), within a web-based application,
backed by a persistent database.

This application can then drive actions within the software-defined infrastructure layer. The app should take actions on behalf of the user, utilising the users' authorisations to the infrastructure layer -
for security and audit reasons the management app should have no authorisations of its own. To provide a strict separation of concerns between operations at the infrastructure level and operations within the deployed infrastructure, we recommend that user accounts with authorisation to deploy and amend infrastructure are kept separate from those used to administer systems deployed within the infrastructure.
The management application provides a convenient view on data stored within that infrastructure, abstracting the complexity of those interfaces. 

Metadata regarding projects, their membership etc, is business data about the projects, (not the research datasets, the subject matter of the research), and its security management processes should be governed by an institution's prevailing business data protection policies, which are not the subject of this paper.

\subsection{Prior Work}
Here, we are going to review some of the existing technologies and tools used to work with sensitive datasets. When we refer to ‘sensitive’ data, we include data that is commercially or government sensitive as well as personal data.

Researchers often create data storage and processing platforms to support their own research, tailored towards the needs of their research fields. Typically, such platforms provide some level of data security but do not limit the researchers’ access to the underlying data, making them unsuitable for working directly with sensitive data from third parties. 

To accommodate this, dataset providers commonly use on-site software to create a de-identified copy of the data which is then transferred to the research institution. For example, a medical imaging research project might utilise widely-used software tools such as: CTP~\cite{ctp} to anonymise the data; XNAT~\cite{marcus2007xnat} to store the anonymised data; and DAX~\cite{harrigan2016van} to run analysis pipelines on the data.
 
Such approaches provide processing flexibility but also bring a number of drawbacks, such as:
\begin{itemize}
\item relying on a high level of confidence in the de-identification process - this may fail if, for example, personal identifiable information appears unexpectedly in free text fields or burnt into image data
\item not addressing the possibility of data re-identification from multiple dataset combination 
\item limiting the ability to audit the entire data processing pipeline as an integrated system (due to the separation of the de-identification and processing systems, which are installed at different institutions) 
\end{itemize}

Additionally, self-hosted systems may be customised to particular environments and designed without reference to scalability and reusability of the software architecture. This makes them unsuitable for supporting the widespread application of rigorous methods to establish secure environments.

\textbf{Secure research environments}

There are presently a number of research environments which offer appropriate levels of security to handle sensitive data. The UK Secure eResearch Platform (UKSeRP), which was developed by the Farr Institute~\cite{jones2016uk} and is now run by Swansea University. The platform allows data owners to upload and set access controls for their data. Significantly for this paper, a new UKSeRP can be instantiated on request, scaled to the needs of individual projects~\cite{SAILdatabankbrochure}. A UKSeRP powers the Dementias Platform UK~\cite{dementiasplatform}, a collaborative secure environment run in a virtual desktop – meaning that data is not transferred to the users, and cannot be erased from the virtual desktop.

The Farr Institute also established the network of Safe Havens used by NHS Scotland, with secure access points to both national and regional havens and remote access granted via VPN to trusted researchers. The havens operate in a federated network, with support and co-ordination provided by a single point of contact: eDRIS (the electronic Data and Research Innovation Service)~\cite{eDRIS}.

Public Good projects can use the UK Data Service’s Secure Lab, which uses a Citrix VPN to allow physical and remote access from within the UK and prevents any download of data~\cite{SecureDataLab}. The Office for National Statistics (ONS) operates a Secure Research Service for approved researchers, with access points within government organisations and in physical ONS ‘Safe Settings’, with plans to increase opportunities for remote access~\cite{SecureResearchService}. Similar secure data schemes are operated by other national governments, such as the German Research Data Centre at the \textit{Institut für Arbeitsmarkt (IAB)}~\cite{IAB} and the French \textit{Centre d'accès sécurisé aux données (CASD)}~\cite{CASD}. These schemes, along with others including the UK Data Service, are part of the International Data Access Network (IDAN)~\cite{IDAN}. Establishing agreements to allow remote access to international data in these schemes is a stated aim of the ONS, making a co-ordinated approach a priority.

SaaS examples of secure data platforms include data.world~\cite{dataworld}, which allows data owners to upload data and maintain strict control over access levels governing the security of work in the platform. Analytixagility by Aridhia~\cite{aridhia} similarly provides a collaborative environment with a role-based permissions model and managed data export.  

Additionally, many big data platforms based on Hadoop technology make use of frameworks that allow granular access to monitor and control data~\cite{ranger}. Products such as SDX by Cloudera~\cite{sdx} offer an integrated environment that assures data governance and security, including the capability to suppress sensitive data~\cite{dataredaction}. Microsoft Azure utilizes Role-Based Access Control (RBAC) and has published a blueprint for organisations to manipulate sensitive data in compliance with American HIPAA and HITRUST regulations~\cite{HIPAA}.

There are also platforms to support collaborative research which don’t have the full functionality of secure research environments, but which conform to similar standards. For example, the open source DataSHIELD software allows researchers to analyse data without it leaving its home environment, offering a workaround enabling collaborative research on sensitive data~\cite{wilson2017datashield}. The Beacon Project established by ELIXIR and the Global Alliance for Genomics and Health (GA4GH) has developed a sharing platform which allows secure data upload and enables users to make queries of datasets and receive yes/no responses~\cite{Beacon}. The establishment of a tiered access system paves the way for this to become a secure research environment in the future.

\textbf{Existing frameworks and standards}

All of the solutions described comply with ISOs 27001~\cite{ISO27001} and 27002~\cite{ISO27002}, and those which handle personal data in the EU must meet the requirements of the EU General Data Protection Regulation (GDPR)~\cite{GDPR}. 
Platforms which handle NHS data must meet the ten high-level security standards set by the National Data Guardian for Health and Social Care \cite{DataGuardian}. Compliance against these standards can be checked using the Data Security and Protection Toolkit published by NHS Digital \cite{DSPToolkit}.

Safe Havens within NHS Scotland’s federated network must comply with the Scottish Government’s Safe Havens Charter~\cite{SafeHavensCharter}. As well as stipulations for operating and hosting a Safe Haven, the charter sets out some technical requirements such as the need for 2fa and an audit log for secure access points, but doesn't provide technical guidance on the establishment of secure environments. The charter in turn is informed by the Guiding Principles for Data Linkage published by the Scottish Government~\cite{DataLinkage}.

The Global Alliance for Genomics and Health (GA4GH) maintains a suite of technical standards for sharing genomic and health related data~\cite{GenomicDataToolkit}; this includes the API for the Beacon project discussed above. These fit within their Framework for Responsible Sharing of Genomic and Health-Related Data~\cite{GenomicSharingFramework}.

A recent addition to the GA4GH standards is Data Use Ontology (DUO)~\cite{DUO}, a system to standardize the terms used to classify health data according to their use and therefore restrictions on usage~\cite{dyke2016consent}.  This has similar aims to the DataTags system developed by Harvard University~\cite{Datatags}. Both systems for data tagging would map to a tiered model of environments for data access.

While there are a number of frameworks and standards for sharing secure data, we were unable to find detailed technical guidance on establishing secure research environments, particularly with regards to treatment of data ingress and egress at different tiers. 

Finally, careful considerations are also required when analysis outputs are extracted from secure environments prior to subsequent dissemination in academic or official publications. Practical recommendations have been provided by Eurostat — the statistical office of the European Union~\cite{eurostat}. These cover both descriptive statistics (such as frequency tables) as well as the output of standard statistical models (such as linear regression). However, to the best of our knowledge, there are currently no guidelines for output checking in the context of more complex statistical and machine learning methods, such as random forests.

The book ~\textit{The Anonymisation Decision-Making Framework}~\cite{elliot2016anonymisation} and article ~\textit{Functional Anonymisation: Personal Data and the Data Environment}~\cite{elliot2018functional} provide very useful frameworks to think about anonymisation and pseudonymisation.

\section{A model for secure data research projects}

\begin{figure*}[htbp]
  \centering
  \includegraphics[width=18cm]{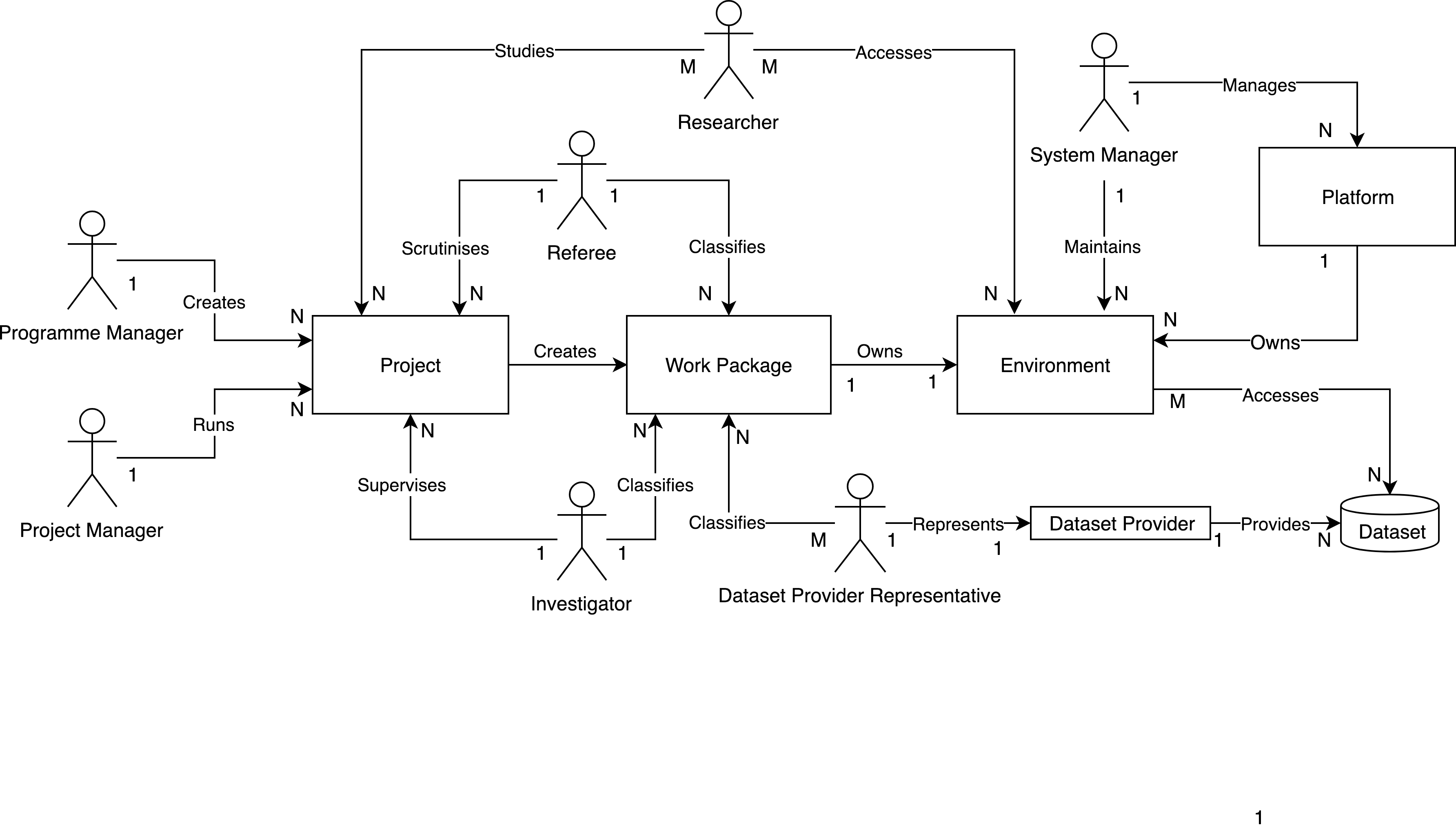}
  \caption{\label{metamodel}  Entity relationship diagram for secure data research.}
\end{figure*}

Figure~\ref{metamodel} describes our model for secure data research.

\subsection{Overview}

\subsection{Work Packages and Projects}
\label{sec:workpackages}

Assessing the sensitivity of a dataset requires an understanding of the base sensitivity of the information contained in the dataset and of the impact on that base sensitivity of the operations that it will undergo in the research project. 
The classification exercise therefore relates to each stage of a project and not simply to the datasets as they are introduced into it.

Classification to a tier is therefore not a property of a dataset, because a particular dataset's sensitivity depends on the data it can be combined with, and the use to which it is put. Nor is it a constant across a project - many projects will begin with identifiable data, and then anonymise it, for example.

In our model, projects are divided into \textbf{work packages}, which we use here to refer to the activities carried out  within a distinct phase of work carried out as part of a project, with a specific outcome in mind. A work package can make use of one or more datasets, and includes an idea of the analysis which the research team intends to carry out, the potential outputs they are expecting, and the tools they intend to use - all important factors affecting the data sensitivity.

Classification should be carried out on the work packages rather than individual datasets. At their first interaction with a Dataset Provider, the research team may only have some idea of how they intend to use the dataset, so we outline the procedures for initial classification of individual datasets in section \ref{sec:initial}.

When a full classification is carried out, Dataset Providers should have a full understanding of any datasets that will be combined and what analysis is intended. This constitutes a work package, even if only one dataset is being classified. If multiple datasets are combined in one work package, a representative from each Dataset Provider will need to agree the classification.

For more information on the classification process, see section \ref{sec:classification} below.

\subsection{Environments and Platforms}

Once a work package has been classified, an appropriate secure research environment is instantiated depending on the tier assigned. 

A project will create one environment for each work package, and will create work packages as required, corresponding to the stages of the project and the current tasks in operation.

Depending on the classification assigned, a work package may instantiate its environment on one of several Platforms. For example, the management system may allow instantiation of environments both on the public cloud or an on-premises private infrastructure, or on a remote HPC cluster. Each of these Platforms may host an environment, and each should be integrated appropriately with the management framework.

\subsection{Datasets}

In our model, datasets only have sensitivity assigned as part of work packages.  Even one dataset on its own should be classified as a work package with the proposed analysis and potential outputs taken into account. A dataset can be analysed in multiple environments if it is included in multiple work packages, but each work package must be classified separately.

When a second project makes use of a dataset already used by a previous project, it is part of a new work package which must be separately classified from scratch. When a project ends, after egress of any reclassified data outputs, its datasets are deleted.

This illustrates our choice to focus on the design of environments for research projects, not a persistent data facility or data trust~\cite{datatrust}.
In future, the relation between environments and datasets must become many-many, to allow for data facilities.

\subsection{Researcher}

A project member, who analyses data to produce results. We reserve the capitalised term ``Researcher'' for this role
in our user model. We use the lower case term when considering the population of researchers more widely.

\subsection{Investigator}

The research project lead, this individual is responsible for ensuring that the project staff comply with
the environment's security policies. Although multiple Researchers will take part in the project, we recommend that a single lead Investigator is accountable to the research institution for a project. Multiple collaborating institutions may have
their own lead academic staff, and academic staff might delegate to a Researcher the leadership as far as interaction with the environment is concerned.
In both cases, the term Investigator here is independent of this - regardless of academic status or institutional collaboration, this individual accepts moral responsibility for the conduct of the project and its members. (Although legally the research institution retains liability.)

\subsection{Referee}

A Referee volunteers to review code or data as it brought into or out of an environment, providing evidence to the Investigator and Dataset Provider Representative that the research team is complying with data handling practices. 

Referees also play a vital role in classifying work packages at Tiers 2 and above. This is particularly important when output data from an analysis environment is classified as a new work package at a different tier.

The importance of an independent Referee for sensitive data handling is discussed further in section~\ref{sec:review}.

\subsection{Dataset Provider and Representative}

The Dataset Provider is the organisation which provided a dataset under analysis. The Dataset Provider will designate a single representative contact to liaise with the research institution.

This individual is the Dataset Provider Representative.

They are authorised to act on behalf of the Dataset Provider with respect to the dataset and must be in a position to certify that the Dataset Provider is authorised  to share the dataset with the research institute, and know whether any further GDPR consent~\cite{GDPR} or other permissions grants are required.

The web management interface will be used to manage any transitions of this responsibility between personnel at the Dataset Provider.

Dataset Provider Representatives must be given accounts within the research institute's system to monitor the use to which their datasets
are put and the process of data classification, ingress, and egress.

There may be additional people at the Dataset Provider who will have input in discussions around data sharing and data classification.
It is the duty of the Dataset Provider Representative to manage this set of stakeholders at the Dataset Provider.

When a work package is classified, this may include multiple datasets, which may come from different providers. When this is the case, each Dataset Provider will require its own Dataset Provider Representative. 

Each of these Dataset Provider Representatives should be involved in classifying the work package that their dataset is used in. They will need to have not only an understanding of the content of their dataset, but an understanding of the datasets it will be combined with, and the analysis the Researchers intend to carry out. 

All Dataset Provider Representatives whose data forms part of a work package will need to come to the same consensus on the tier used to analyse that work package in order for the project to proceed.

\subsection{Programme Manager}

A designated staff member in the research institution who is responsible for creating and monitoring projects and environments and overseeing a portfolio of projects.
This should be a member of professional staff with oversight for data handling in one or more research domains.

The Programme Manager can add new users to the system, and assign users to specific projects. They assign Project Managers and can, if they wish, take on this role themselves.

\subsection{Project Manager}
A staff member with responsibility for running a particular project. This role could be filled by the Programme Manager, or a different nominated member of staff within the research institution.

While the Programme Manager should maintain responsibility for adding users to the user list, and can add users to projects, the Project Manager should also have the authority to assign existing users to their project. To do this they will need to be able to view and search existing users.

\subsection{System Manager}

A member of staff responsible for configuration and maintenance of the secure research environment. They also manage the secure platforms on which environments are instantiated.

\section{Environment Tiers}

Our recommendation for secure information processing tiers is based on work which has gone before. We have begun with the UK government classifications~\cite{classifications} as a base, and reconciled these to the definitions of personal data, whether or not something is 'special category' under the GDPR~\cite{GDPR} or  relates to criminal convictions, and related them to common activities in the research community. 

Where input datasets contain personal data, consideration should always be given at the outset to minimising the personal data, including by pseudonymisation or anonymisation. 

\textbf{Pseudonymised data} is still personal data, as it can be re-identified by those who hold the key to turn pseudonyms back into individual identifiers. Note that this may include synthetic data derived from personal data, or models trained on personal data, depending on the methods used to synthesise the data or generate the models.

\textbf{Anonymised data}, including pseudonymised data where that key is destroyed, is not personal data when it is impossible to re-identify any living individual from it. However, if the quality of anonymisation is ambiguous or if individuals can be identified when the anonymised data is combined with another dataset, there is still a danger of personal information being revealed. Such data would by definition not be anonymised, and would therefore be personal data. The GDPR~\cite{GDPR} and ICO guidance~\cite{ICOpersonaldata} make this clear. The question as to \emph{whether} such re-identification is possible or not is a very subtle one, and the assessment of this risk is critical to the assignment of security tiers.

We emphasise that data classification is based on considering the sensitivity of all information handled in the project, including information generated by
combining or processing input datasets. In every case, the categorisation does not depend only on the input datasets, but on combining information
with other information or generated results. We refer to the combination of this information as a work package, which is defined in section ~\ref{sec:workpackages} above.

Derived information may be of higher security tier than the information in the input datasets (for example, information on the identities of those who are suspected to possess an un-diagnosed neurological condition on the basis of analysis of public social media data.) This should form part of the information constituting a work package; when a project team believes this will be the case, they should make this clear to the Dataset Provider Representative, and the work package should be classified at the higher tier of secure environment.

If it becomes apparent during the project that intended analysis will produce this effect then the inputs should be treated as a new work package with this extra information, and classified afresh, following the full classification process below.

In the following, ``personal data'' follows the GDPR~\cite{GDPR} definition: information from which a living individual is identified or identifiable. It excludes information about individuals who
are dead.

\subsection{Tier 0}

Tier 0 environments are used to handle publicly available, open information, where all generated and combined
information is also suitable for open handling. 

Tier 0 applies where none of the information
processed, combined or generated includes personal data, commercially sensitive data, or data which will have legal, political or reputational consequences in the event of unauthorised disclosure.

Tier 0 data should be considered ready for publication.

Although this data is open, there are still advantages to handling it through a managed data analysis infrastructure. 

Management of Tier 0 datasets in a visible, well-ordered infrastructure provides confidence to stakeholders as to the handling of more sensitive datasets. 

Although analysis may take place on personal devices, the datasets should still therefore be listed through the inventory and curatorial systems of a managed research data environment.

Finally, audit trails as to the handling of Tier 0 datasets reduce risks associated with misclassification - if data is mistakenly classified as lower tier than it should be, we still retain information as to how it was processed during the period of misclassification.

\subsection{Tier 1}

Tier 1 environments are used to handle, process and generate
data that is intended for eventual publication or that could be published without reputational damage. 

Information is kept private in order to give the research team a competitive advantage, not due to legal data protection requirements.

Both the datasets and the proposed processing must otherwise meet the criteria for Tier 0.

It may be used for anonymised or synthetic information generated from personal data, where one has absolute 
confidence in the quality of anonymisation. This makes the information no longer personal data. This does \textbf{not} include pseudonymised data which can be re-identified in combination with a key or other dataset. Please refer to the definitions above.

It may also be used for commercial data where commercial consequences of disclosure would be no impact or very low impact, with the agreement of all parties.

\subsection{Tier 2}

Tier 2 environments are used to handle, combine or generate information which is not linked to personal data.

It may be used for pseudonymised or synthetic information generated from personal data, where classifiers have strong
confidence in the quality of pseudonymisation. 
The risk of processing it so that individuals are capable of being re-identified must be considered as part of the classification process.

The pseudonymisation or anonymisation process itself, if carried out in the research organisation, should take place in a Tier 3 environment.
See section \ref{sec:egress} for a full discussion of audit of pseudonymisation or anonymisation code.

A typical model for a project handling personal data will be to instantiate both Tier 2 and Tier 3 environments, with pseudonymised or synthetic data generated in 
the Tier 3 environment and then transferred to the Tier 2 environment.

Tier 2 environments are also used to handle, combine or generate information which is confidential, but not, in commercial or national security terms, sensitive.
This includes commercial-in-confidence datasets or intellectual property where the legal, commercial, political and reputational consequences from disclosure are low. Where such consequences are not low, Tier 3 should be used.

At Tier 2, the most significant risks are `workaround breach` and the risk of  mistakenly believing data is anonymised, when in fact re-identification might be possible.

Tier 2 corresponds to the government OFFICIAL classification.

\subsection{Tier 3}

Tier 3 environments are used to handle, combine or generate personal data, other than personal data where there is a risk that disclosure might pose a substantial threat to the personal safety, health or security of the data subjects (which would be Tier 4).

It includes pseudonymised or synthetic information generated from personal data, where the classifier has only weak
confidence in the quality of pseudonymisation.

Tier 3 environments are also used to handle, combine or generate information, including commercial-in-confidence information and intellectual property, which is sensitive in commercial, legal, political or national 
security terms. 
This tier anticipates the need to defend against compromise by attackers with bounded capabilities and resources.
This may include hacktivists, single-issue pressure groups, investigative journalists, competent individual hackers and the majority of criminal individuals and groups.
The threat profile excludes sophisticated, well-resourced and determined threat actors, making targeted attacks, such as highly capable serious organised crime groups and state actors.
This corresponds to the governmental ‘OFFICIAL–SENSITIVE’ descriptor. \cite{classifications}

The difference between Tier 2 and Tier 3 environments is the most significant in this model, both for researcher productivity and organisational risk. 

At Tier 3, the risk of hostile actors attempting to break into the secure environment becomes significant.

\subsection{Tier 4}

Tier 4 environments are used to handle, combine or generate personal data 
where disclosure poses a substantial threat to the personal safety, health or security of the data subjects.

This also includes handling, combining or generating datasets which are sensitive in commercial or national 
security terms, and are likely to be subject to attack by sophisticated, 
well-resourced and determined actors, such as serious organised crime groups and state actors. This
tier corresponds to the UK government `SECRET' categorisation~\cite{classifications}.

It is at Tier 4 that the risk of hostile actors penetrating the project team becomes significant.

\section{Data lifecycle}
\label{sec:datalifecycle}

\subsection{The Classification Process}
\label{sec:classification}

We now describe a recommended process through which research activities with data of any degree of sensitivity are classified. 

We envisage that classification discussions will be ongoing between the research team and Dataset Providers whose data is included in a \textbf{work package} (see section \ref{sec:workpackages}). Rather than a decision taken at a point in time, classification should be viewed as an ongoing question which should be revisited in response to new knowledge about the data, or changes to the plans of the research team.

\begin{figure*}[htbp]
  \centering
  \includegraphics[width=19cm, height=23cm]{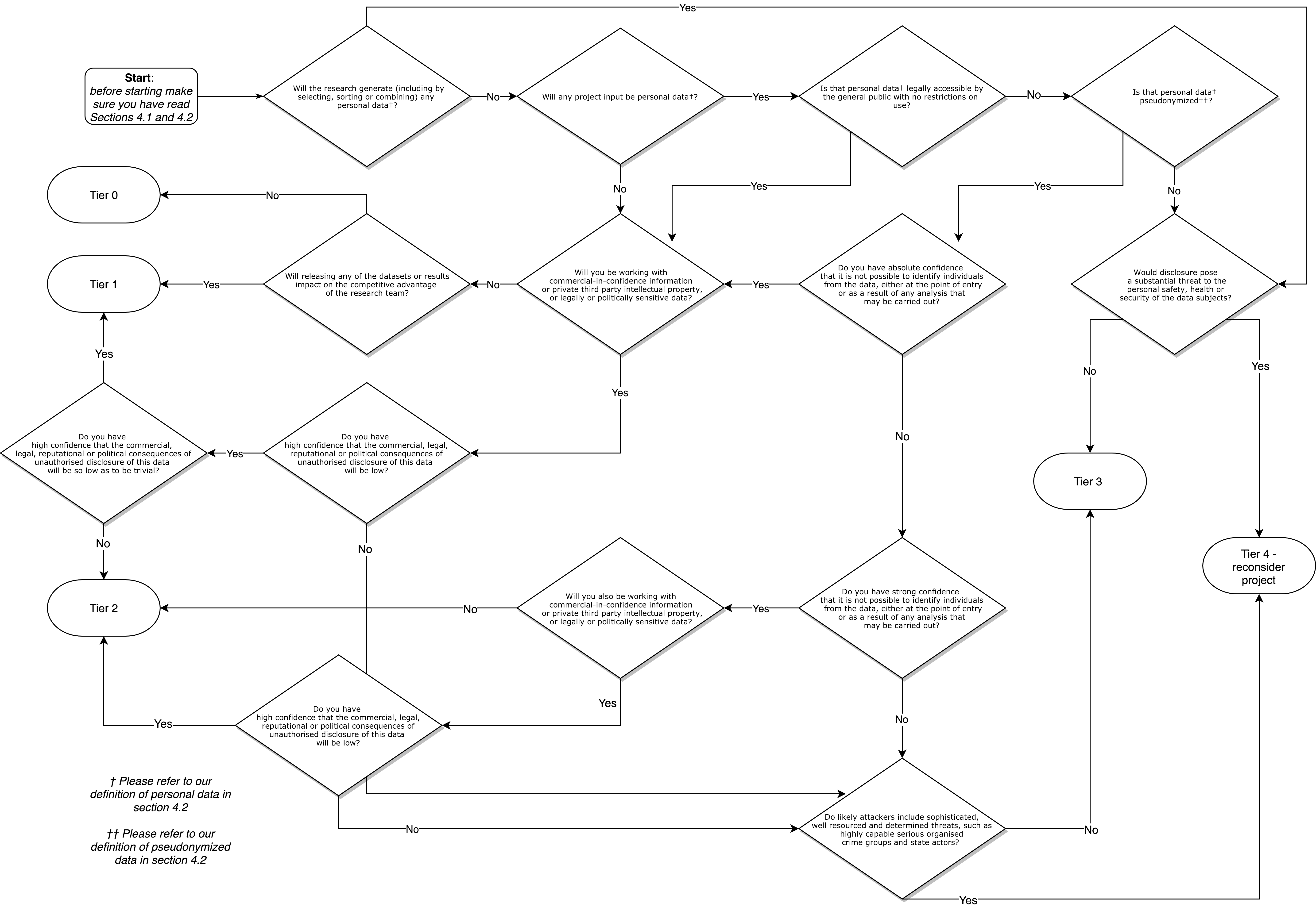}
  \caption{\label{tiers} Flowchart describing questions for data tier selection for our recommendation of five tiers.}
\end{figure*}

That said, there will need to be decision points at which all parties agree that data can be ingressed into an environment at a particular security tier. For this purpose, we have designed a flowchart (Figure \ref{tiers}) which should be used to classify work packages at these points, and which should be embedded in the management framework to record results.

We anticipate that there will be five approximate stages to the data classification process, outlined below:

\begin{itemize}
    \item Dataset Providers and research teams hold initial conversations about the sensitivity of a work package - that is, a set of research activities involving the dataset (section \ref{sec:workpackages}), using the flowchart in Figure \ref{tiers} to guide the conversation.
    \item During these conversations, the work package is given an initial preliminary classification that all parties are confident is equal to or higher than the true classification of the work package.
    \item An environment is then instantiated at the preliminary tier and the dataset ingressed into it to provide the Investigator with the complete information on which to base a final classification.
    \item At some point after this,  all classifiers (the Investigator, Dataset Provider Representative and a Referee if necessary) go through the flowchart in Figure \ref{tiers} and agree on a final classification for the work package. Analysis can only start once classification is agreed.
    \item The dataset will either stay in the initial environment environment or be ingressed into a new environment at a different tier.
    \item If, after this, anything changes such as new knowledge about the dataset, the intended analysis changing, or the research team wanting to ingress a new dataset to combine with the existing one, this creates a new work package which must be classified afresh by all classifiers.
    \item If the classification tier resulting from this process differs from that of the existing environment, the existing dataset and any output of the work package must be transferred to a new environment of the appropriate tier, following the data egress process outlined in section \ref{sec:egress}.
\end{itemize}

\subsubsection{Initial Classification}
\label{sec:initial}

During initial classification conversations, the research team will not have access to the dataset, so their judgment will be limited.

For this reason, we recommend that at some point during these initial  discussions, the dataset is ingressed into a Tier 3 environment to which the Investigator has access, so that they have complete information on which to base full classification.

We recommend a Tier 3 environment as most datasets will be Tier 3 or lower, and ingressing into a Tier 3 environment initially is therefore erring on the side of caution. We recommend that this is done even when the Dataset Provider believes the data should be classified at Tier 0, Tier 1 or Tier 2, as the risk of ove-classifying initially before instantiating a lower-tier environment are lower than the risks of under-classifying at this stage.

The exception to this should be where \textit{any} party feels there is a possibility that data may be Tier 4 (see Figure~\ref{tiers}). In this case, all parties should reconsider whether to proceed with the project, and the research team should work with the Dataset Provider to establish the legal sensitivity of the data before performing any ingress.

\subsubsection{Full Classification Process}

After the initial ingress of data to a Tier 3 environment, the Investigator and the Dataset Provider Representative should carry out full classification before proceeding with the project. 

A central premise which defines our approach to classification is that datasets should not be classified in isolation but as part of a \textbf{work package}. This is a term we use to refer to the datasets and activities involved in a distinct phase of work carried out as part of a project, with a specific outcome in mind. 

A work package can be made up of one or more datasets, and includes, in addition to the data, an idea of the analysis which the research team intends to carry out, the potential outputs they are expecting, and the tools they intend to use - all important factors affecting the sensitivity of the data.

If the research team only intends to analyse one dataset, they should communicate to the Dataset Provider how they intend to analyse the data and what outputs they are expecting. With knowledge of this intended use, the Dataset Provider Representative should assess which classification tier they recommend for the data by answering the questions set out in the flowchart below (Figure~\ref{tiers}). The project Investigator should separately go through this process once they have access to the data in a Tier 3 environment. If the data is likely to be Tier 2 or higher, we additionally recommend appointing an independent Referee to provide their opinion (See~\ref{sec:review}).

The project should only proceed if the Investigator, the Dataset Provider Representative, and the Referee if applicable, can come to a consensus for which tier the dataset should be classified at. Where disagreements occur over which classification tier to apply, the management system should notify the users, who should attempt to reach consensus.

If consensus cannot be reached, the Investigator should seriously reconsider whether to proceed with the project - the risks associated with projects where there is not consensus as to data handling between the Dataset Provider and research team are significant.
In the  event that the project will proceed in the absence of consensus over classification, the highest tier proposed should apply. Alternatively, the Dataset Provider may wish for their dataset to be removed from the system if consensus cannot be reached - if this is the case, the storage volumes (see section~\ref{sec:storage}) should be deleted.

Once the work package has been classified, a new environment should be instantiated at the appropriate tier, unless it has been classified at Tier 3, in which case analysis can begin in the existing environment. In either case, no analysis should begin until all classifiers have agreed on the classification, and this agreement is recorded in the system. 

If, at any point during the project, the research team decides to analyse the data differently or for a different purpose than previously agreed, this constitutes a new work package, and should be newly classified by repeating this process. This is also the case if the team wishes to ingress another dataset in combination with the existing one, specific recommendations for which are laid out below. 

\subsubsection{Combining Multiple Datasets}

Researchers may want to combine two or more datasets and analyse them together in the same environment. It is important that any classification of a dataset takes into account the data it will be used in combination with, as this could change the sensitivity by, for example, making data easier to de-anonymize. For this reason, we lay out specific recommendations for classifying datasets in combination below.

If the research team knows that they will analyse datasets in combination, they should begin by ingressing them separately into their own Tier 3 environments. 

The work package on which full classification is carried out will consist of each dataset they intend to combine, in addition to information about the intended analysis and expected results. The research team should communicate to each Dataset Provider as much detail as is appropriate about the other datasets their data will be combined with. 

A Dataset Provider Representative from \textit{each} of the combined datasets should then separately go through the classification questions in Figure~\ref{tiers}, as well as the Investigator and a Referee if necessary. Consensus from each party, including all Dataset Provider Representatives, is needed before moving on. Once consensus has been reached, a new environment should be established at the agreed tier and the datasets ingressed into it.

To combine a new dataset with a work package that has already been classified, the process followed should be substantially the same, as this would create a new work package separate to the one that had previously been classified. As above, the research team should communicate to each Dataset Provider detail about the other dataset(s) their data will be in combination with, and go through the full classification process from the start. If the classification tier decided upon, taking into account new data added, is the same as the existing work package, the new dataset may be added to the existing environment. If not, this is a new work package, and a new environment should be instantiated to carry out the combined analysis.

If the research team intends to combine multiple datasets in future, but wish to begin analysis on one dataset, they should include details about any intended future combination as part of the initial work package, and this should be taken into account when classifying. In this case, it is possible that the research team could introduce a previously agreed dataset at a later stage without revisiting the classification process. However, even in this case, we recommend that new datasets go through the classification process, in case the Dataset Provider's position has changed.

\subsubsection{Data egress and new classification}
\label{sec:reclassification}

The initial classification of a work package will often be for the purpose of ingress into an initial high-tier environment to carry out anonymisation, pseudonymisation or synthetic data generation work, with the intention of making the data appropriate for treatment in a lower-tier environment. 

Researchers may equally want to conduct analysis on output data from a previously analysed dataset if they believe it would be useful to continue the research.

It is a central premise of our model that any output data is classified as a new work package, separate from the work package that it is derived from.

Any data which is intended to be removed from an analysis environment results in the creation of a new work package and the full classification process should be followed, whether for the purposes of publishing output data (in which case the output data must be Tier 1 or Tier 0), or analysing it in a new environment. As with the initial classification, the Dataset Provider Representatives for each dataset included in the work package need to agree with the classification, along with the Investigator.

As a convenience, if the resulting derived data is in a form which has been agreed with Dataset Provider Representatives at the initial classification stage – for example, a summary statistic which was the intended output for analysis of a sensitive dataset - then this re-classification may be pre-approved.

In this case, Investigators may consult an independent Referee and classify the work package without returning to the Dataset Provider Representative, if they have high confidence that the outputs are as anticipated during the initial classification.

We recommend that, whenever data egress is conducted with the intent of establishing a new environment for further research, a Referee is consulted to ensure balance (see Section~\ref{sec:review}).

In all cases, classification of a work package at the point of egress should be done with all parties fully aware of the analytical processes which created the derived data from the initial work package – and these should be fully reproducible.

In most cases, this will necessitate an executable script describing, in code, the processes used. The authors do not believe that a spreadsheet can be properly audited for this. This script should be provided so that those classifying the output data can understand and verify how the output data was generated from the more sensitive work package dataset.

After the full classification process has been followed, a new environment can be created with the former volume used for data egress now mounted as a new secure data volume within a new environment, at a different tier (see section~\ref{sec:storage}). The existence of this environment as a ''derived environment'' should be noted, with the originating environment's ID and the analysis script preserved.

Common examples here are the classification of a neural network trained on sensitive data, or the development of a statistically similar synthetic dataset. The question as to whether these can leak personal information from the training set is a very subtle one: hence our recommendation of expert peer review and consultation with the original Dataset Provider(s).

Similar considerations apply to the creation of derived datasets by sampling from a larger dataset - while not of lower GDPR sensitivity, these can often be of lower commercial sensitivity, justifying a different tier. The same classification process should be followed as above and noted on the original DPIA.

\subsubsection{Publishing or returning data}
\label{sec:egress}

If it is agreed that the output data work package can be classified as Tier 1 or Tier 0 and published, an appropriate environment at this tier should be created, and data can be copied out directly via Secure Copy Protocol (SCP).

As an exceptional process, it may be necessary to release generated
sensitive data back to the original Dataset Provider, or to transfer to an
independent secure data environment.

In this circumstance, then the following process should be followed:

The Dataset Provider Representative and Programme Manager should authorise this in the management system,
which makes record of the fact. The expected IP range and time duration for the extraction should be specified at this time. This should temporarily make available a new secure volume accessible outside the environment using 
the credentials of the Dataset Provider Representative, from which data may be copied out.

\subsubsection{Personal data handling}
\label{sec:personaldata}

If personal data (see definition in section~\ref{sec:definitions} below) is handled or generated, the Programme Manager and the Project Manager must verify that 
there is a fair, transparent and lawful basis both for the Dataset Provider to share the information with the research institution and for the research institution to use that information as planned in the project. If the personal data were originally collected on the basis of consent, fresh consent for use in the project will usually be required. Note that this is a separate consent specifically for the purposes of ensuring GDPR compliance and should not be equated with any ethical consent that may be necessary. 
If the personal data were originally collected on a lawful basis other than consent, it will almost always be possible to use them in a research project provided that adequate safeguards are in place. A Data Protection Impact Assessment (DPIA) must be carried out and the final version recorded in the management system, and
in data protection processing records of all relevant organisations.

The use of personal data that is available on social media or other websites is governed by the terms and conditions of the relevant website and also by the laws of confidence, privacy and data protection. It must not be assumed to be free to use for a particular project without checking.

The project may need to be approved by an Ethics Advisory Group before authorisation - the completion of this process should also be recorded in the management framework by the Project Manager.

\subsubsection{Data Sharing Agreement}

Any contractual terms regarding the datasets should be added to the metadata for all datasets which have been classified, including any additional commitments to be made by Researchers on joining the project.

A template agreement with Dataset Providers normally used by the research institution (ideally, already existing) should be generated by the system before a dataset is ingressed to the initial Tier 3 environment. This should be a formal data sharing agreement 
as required under data protection law, and drafted with the benefit of legal advice.

After the full classification process is carried out, the agreement may need to be updated. We recommend that the wording of the initial agreement is such that it enables research at the tier anticipated by the Dataset Provider, or is worded flexibly enough that changes after the initial classification are minimal.

The executed agreement itself, based on the template, should be physically or digitally signed and stored within a secure document volume (see section~\ref{sec:storage}), with an index into it maintained within the database of the web management tool. A copy of the agreement should be automatically stored in the same location as the dataset for user reference when working with the data in the analysis environment.

\subsection{How to classify work packages}
\textbf{Figure~\ref{tiers}} below shows a flowchart with the questions that should be used to allocate a work package to one of the security tiers.

These questions should be implemented as a series of dialogues within the management system.

Each person who is classifying the data should \textbf{separately} go through this flowchart and make their own classification – these are the Investigator, the Dataset Provider Representative, and, if the data might be Tier 2 or above, an independent Referee [see section \ref{sec:classification} above]. 

If the Investigator and the Dataset Provider Representative disagree about the tier classification, they should find a consensus before starting the project. This is particularly true in the case of commercial-in-confidence information, which should be treated as Tier 2 unless the Dataset Provider Representative agrees it can be lower.

Please note that the flowchart should only be used to assess into what tier the work package should fall. It does not include other actions researchers should take when dealing with personal data, such as considering ways to minimise the personal data they need to use.

\subsubsection{Flowchart definitions}
\label{sec:definitions}

In the flowchart, the terms below are defined as follows:

\textbf{Personal data} is any information relating to an identified or identifiable \textbf{living individual} (see below); an 'identifiable' living individual is one who can be identified, directly or indirectly, in particular by reference to an identifier such as a name, an identification number, location data, an online identifier or to one or more factors specific to the physical, physiological, genetic, mental, economic, cultural or social identity of that natural person.

The term 'indirectly' here indicates that this includes data where identification is made possible by combining one or more sets of data, including synthetic data or trained models.
\textbf{See also section \ref{sec:personaldata} above.}

A \textbf{Living individual} is an individual for whom you do not have reasonable evidence that they are deceased. If you’re unsure if the data subject is alive or dead, assume they have a lifespan of 100 years and act accordingly. If you’re unsure of their age, assume 16 for any adult and 0 for any child, unless you have contextual evidence that allows you to make a reasonable assumption otherwise~\cite{nationalarchives2018}. 

\textbf{Pseudonymised data} is personal data that has been processed in such a manner that it can no longer be attributed to a specific living individual without the use of additional information, which is kept separately and subject to technical and organisational measures that ensure that the personal data is not attributed to an identified or identifiable living individual. 

Two important things to note are that pseudonymised data:
\begin{itemize}
    \item \textit{is still personal data. It becomes anonymised data, and is no longer personal data, only if \emph{both} the key data connecting pseudonyms to real numbers is securely destroyed, \emph{and} no other data exists in the world which could be used statistically to re-identify individuals from the data.}
    \item depending on the method used, it often \textit{includes} synthetic data and models that have been trained on personal data. Expert review is needed to determine the degree to which such datasets could allow individuals to be identified. Such data should be treated as pseudonymised data until such expert review is received.
\end{itemize}

It is important that both researchers and Dataset Providers consider the level of confidence they have in the likelihood of identifying individuals from data. 

\textbf{Anonymised data} is data which under no circumstances can be used to identify an individual, and this is less common than many realise~\cite{Rocher2019}.

Our model specifies three levels of confidence that classifiers can have about the likelihood of re-identification, with each pointing to a different tier - absolute confidence, where no doubt is involved, strong confidence, or weak confidence. Classifiers should give sufficient thought to this question to ensure they are classifying data to the appropriate sensitivity. 

\textbf{Commercial-in-confidence data} is information which, if disclosed, may result in damage to a party’s commercial interest, intellectual property, or trade secrets.

\subsection{Data ingress}

The policies defined here minimise the number of people who have access to restricted information before it is in the secure research environment.

When a work package is fully classified, data will either be ingressed from the initial Tier 3 environment to a new analysis environment, or will stay in the original Tier 3 environment, depending on sensitivity. The web management workflows should ensure that all parties have reached consensus on the classification tier at this stage before allowing analysis to begin. 

Initially, datasets should be transferred from the Dataset Provider into a Tier 3 research environment once a Data Sharing Agreement has been signed, and the executed agreement is within the system. The workflows in the web management framework should enforce the need for this agreement to be reached before transfer. 

The transfer process should be initiated by the Project Manager in the management framework, opening a new empty secure data volume (see section~\ref{sec:storage}) for deposit.

The Investigator should authorise the mounting of the data volume in the Tier 3 environment, using the web interface.

After their dataset has been transferred, the Dataset Provider Representative should immediately use the management framework to indicate that the transfer is complete. In doing so, they lose access to the data volume. 

Data should always be uploaded directly into the secure volume
to avoid the risk of individuals unintentionally retaining the dataset for longer than intended.

While open for deposit, this volume provides an additional risk of a third party accessing the dataset. We recommend that to deposit the dataset, a time limited or one-time access token, providing write-only access to the secure transfer volume, should be granted and transferred via a secure channel to the Dataset Provider Representative.

The Project Manager may remind the Dataset Provider Representative that this has been activated by a phone call or email, but the details for access must only be provided in the management framework, not over other channels, to ensure only authorised personnel have access.

\subsubsection{Non-internet data ingress.}

If the Dataset Providers' network bandwidth to the cloud service is significantly worse than the research institutions, or the data has not yet been digitised,
then the data may need to be provided to the research institution via a physical medium. Protocols in this case are beyond the scope of this paper.

\subsubsection{Verification of integrity of transfer}

The management system should make available, once the transfer is complete, a hash of the deposited dataset. The Dataset Provider Representative 
should then confirm to the system that the hash matches a hash generated by the Dataset Provider. If this is not the case, the Programme Manager and the Project Manager must be alerted. This integrity check may potentially be automated given the appropriate choice of upload client and endpoint.

The management system should regularly automatically re-verify the integrity of the dataset, and alert Programme and Project Managers if the integrity is not maintained.

\subsection{Data deletion and project closure}

On completion of a project, following completion of the reclassification or egress process for any
data or publications generated, including those required by funders for long-term
storage, the storage volumes (see section~\ref{sec:storage}) for all associated environments should be deleted. (Archival preservation should
be the subject of a different system with different requirements.)

The project itself should remain in the system with metadata indicating its completion, and the list of
environments and data volumes created retained.

\section{The Analysis Environment}
\label{sec:environment}

\subsection{Software library distributions}
\label{mirrors}

Maintainers of shared research computing environments face a difficult challenge in keeping research algorithm libraries and platforms
up to date - and in many cases these conflict. While sophisticated tools to help with this such as Spack~\cite{spack} exist,
the use of single-project virtual environments opens another possibility: downloading the software as needed for the project
from package managers such as the Python package index.

For Tier 1 and 0 environments, packages can be installed directly from the appropriate official package server on the public internet. However, to support on-demand package installation in Tier 2 and above environments, without access to the external internet, requires maintenance of mirrors of the official package repositories inside the environment.

In addition to making a much wider range of packages available for some tiers, the use of package mirrors inside the environment means that the set of default pre-installed packages on analysis machines can be kept
to a minimum, reducing the likelihood of encountering package-conflict problems and saving on
System Manager time.

Malicious software has occasionally been able to become an official download on official package mirror, including the inclusion of addition of malicious code as accepted contributions to existing open source projects.
Defense in depth means this is a low risk, since the environment will not have access to the internet, but must still be guarded against.

For Tier 2, we recommend maintaining a full mirror of the official package server. To guard against "zero day" exploits of popular packages, a short delay may be considered between a package being updated on the official package server and being updated on the internal mirror. However, extended delays in propagating updates to the internal mirror risks security fixes remaining unavailable long after they have been released on the official mirror. An extended delay in updating the internal mirror also diminishes the usefulness of the packages available, as researchers are often early adopters of new packages and features.

For Tier 3, we recommend minimising the threat surface from potentially malicious packages by only mirroring a white-listed subset of the full package lists from the official servers. It is feasible to perform a full independent code review of all white-listed packages at this threat level. Instead, we recommend this white-list be based on easily auditable properties of packages, such as the size of the developer and user communities, the availability of source code, and the frequency of updates. We are working on a future publication of an automated model for white-listing packages on the basis of information that can be mined from package metadata and commit records.

\subsection{Internal web services}

Using the web browser and in programmatic connections, users inside the environments should be able to connect to
local, protected instances of key internet services for research, such as:

\begin{itemize}
    \item Git, with service mirror of a git version control management tool such as Gitlab
    \item Collaborative paper-authoring tool such as HackMD or Overleaf
    \item A local relational database such as PostgreSQL
\end{itemize}

\subsection{Languages and tools}

The environment must include the core tools of a professional data science programming environment.
It would not be fruitful to include the full list here, and such a list would quickly age.
However, we highlight critical elements of such an environment below:

\begin{itemize}
    \item The R analysis environment \cite{RCoreTeam}, with package mirrors of CRAN and Bioconductor
    \item Python, with package mirrors of PyPI and Conda
     \item C, C++, and Fortran compilers, with a distribution mirror for installing common packages, such as apt, yum or chocolatey.
\end{itemize}

The following are also important, but less so than the above to our user community:

\begin{itemize}
    \item The Julia programming language and its packaging system
    \item A JVM programming environment, with Java, Scala, and Groovy and a mirror of Maven Central.
    \item .NET Core and Mono
\end{itemize}

Critical to the environment are the tools we use to develop analyses:

\begin{itemize}
    \item Programming text editors, such as vim, emacs, atom and visual studio code
    \item Multi-language notebook programming interface, such as Jupyter
    \item RStudio
    \item Command line scripting shell such as bash, with access to the installed programming languages.
\end{itemize}
    
We note that the  data scientists within the communities we are targeting tend to prefer programming tools over Excel, SPSS or Stata, so these do not form part of our recommended list.

Although we recommend a minimal pre-installed list, with most research software installed from package mirrors, machine learning toolboxes such as PyTorch and Tensorflow
can be problematic to configure oneself, so pre-installing compiled instances of these tools these are worthwhile.

\subsection{At-scale computing}

The environment must be able to scale to handle large and complex datasets, so, within the management framework for the environment, it must be possible
to instantiate multiple compute nodes, and small clusters.

Software must also be made available to access these using both high-throughput and high-performance programming tools, (at time of writing, including MPI and Spark). 

Multi- and many-core computing is also vital to modern data science, so the environment must make available
many-core computing chips and the libraries to
enable this, such as cuDNN and OpenMP.

\subsection{Storage}
\label{sec:storage}

What storage volumes should exist in the analysis environment?

A \textbf{Secure Data Volume} is a read-only volume that contains the secure data for use in analyses. It is mounted read-only
in the analysis environments that must access it. One or more such volumes will be mounted depending on how many managed secure datasets the environment has access to.

A \textbf{Secure Document Volume} contains electronically signed copies of agreements between the Dataset Provider and the
research institution.

A \textbf{Secure Scratch volume} is a read-write volume used for data analysis. Its contents must be automatically and regularly deleted. Users can clean and transform the sensitive data with their analysis scripts, and store the transformed data here.

An \textbf{Output volume} is a read-write area intended for the extraction of results, such as figures for publication. See~\ref{sec:egress} below.

The \textbf{Software volume} is a read-only area which contains software used for analysis. 

A \textbf{Home volume} is a smaller read-write volume used for local programming and configuration files. It should not be used for data analysis outputs, though this is enforced only in policy, not technically. Configuration files for software in the software volume should point to the Home volume.

\section{The Management Framework}

If badly constructed, the web-based management framework described here might have access to make changes to all the secure data volumes, and the identity management system (such as Active Directory) which defines which users can access which datasets.

Compromise of this web application would therefore be a serious security risk. We do not re-discuss here best practices for secure web programming: we assume implementers of this policy framework will follow these, and will, in particular, commission penetration testing and security audit of the framework.

However, we discuss now some aspects of how this web application should be designed in order to maintain security.

The management application should access the software defined infrastructure provision through the user credentials of the user logged into it, rather than a separate 'service user' identity. It forms only a convenient interface to powers that those same credentials could have wielded if logged into the platform's management portals directly. Credentials are not stored with the application server, but rather, proxied from the logged in user's identity.
For this integration to be possible in a secure fashion, the platform must expose its management capabilities through APIs secured through the same login mechanism as used to manipulate them through its own management portals. 

The primary management application, with its database containing the IDs of the secure data volumes,
should be accessible only within the secure network, meaning it can be accessed only from managed devices (see section~\ref{devices}).

However, certain processes need to be able to be carried out through the management framework outside that network, such as the ingress process carried out by a Dataset Provider Representative, or the egress workflow for lower tier environments. Therefore, a subset of views of this application, only, may be made accessible via a restricted proxy, outside the secure network. 

For example, when the IP address range for an upload process for data ingress is entered by the Programme or Project Manager, the appropriate page containing information for the Dataset Provider Representative will be temporarily made available outside the secure network, with firewall rules automatically set appropriately for the duration of the ingress process.

\section{Access to the system}

\subsection{User Devices}
\label{devices}

What devices should Researchers use to connect to the secure research environment?

Should a research institute restrict the access to install and configure software on its researchers' computers?

We note that this is perhaps one of the most controversial and sensitive areas for research organisations.
Scientists and developers defend the devices on which they spend their working lives with reasonable jealousy, and any security measures which cut into researcher productivity or autonomy, or are perceived to, need to be carefully defended.

We define two types of devices: 

\begin{itemize}
    \item Managed devices
    \item Open devices
\end{itemize} 

It is \emph{not} recommended that either kind of 
devices should be used to analyse datasets containing sensitive information
- for handling sensitive data, researchers should use these devices to connect to secure remote analysis environments, as discussed below.

\subsubsection{Managed Devices}

Managed devices must, by default, when issued, not have administrator/root access.

\emph{However}, it is unlikely to be effective from either a cost or productivity perspective to provide researchers with a managed device \emph{solely} for the purposes of connecting to secure research environments, so the capability for software-based research on these devices is therefore also needed.

They should have an extensive suite of research software installed.

This should include the ability to install packages for standard programming environments without use of root (such as \verb|pip install, brew install.)|

Researchers should be able to compile and run executables they code in \textbf{user space.}

The setup should include a hypervisor for managing VMs for activities that need root/administrator access.

\subsubsection{Open Devices}

Staff researchers and students should be able to choose that an employer-supplied device should instead have an administrator/root account to which they do have access.
These devices are needed by researchers who work on a variety of bare-metal programming tasks.

However, such devices must not be able to access Tier 3 or 4 environments.

They may include personal devices such as researcher-owned laptops.

When affordable, researchers may wish to choose a Managed device for their desktop machine and an Open device for their laptop.

\subsection{Connections to the secure environment}

How should Researchers connect to the remote analysis environment?

\subsubsection{Secure remote desktop}
A secure remote desktop connection, allowing access to graphical interface applications, should be provided to allow researchers to connect to the remote secure research environment. 

Our Azure reference implementation provides remote desktop access via access nodes running Microsoft's Remote Desktop Services (RDS) web application, with mandatory two-factor authentication. RDS permits functionality such as disk and clipboard sharing between local and remote machines to be configured on a per environment basis. This allows different controls on data ingress and egress via the access nodes to be set for different tiers.  Alternative platforms such as Apache Guacamole provide equivalent functionality.

For Tier 2 and above, our RDS access nodes are the only parts of the secure research environment open to inbound connections from outside the environment and do not permit any disk, clipboard or device sharing between local and remote machines.

\subsubsection{Secure shell}

Another key question here is the use of ``secure shell'' connections.

Such text-based access is sufficient for some professional data scientists, with the provenance
information provided by command-driven data analysis a primary driver for this preference (processes can 
easily be reproduced based on the commands typed.) 
When not needed, providing a remote desktop interface adds complexity and therefore risk.

However, command line access typically provides very convenient ways for extraction of data files from remote environments (\verb|scp, rsync|), and for all kinds of additional services to be 'tunnelled' through the SSH connection, opening many security holes. While we can forbid these by policy, their convenience makes workaround breach likely, so they must be blocked technologically.

How, therefore, can we provide the command line user experience for access to remote secure computing environments while maintaining security? Casual, unthinking copy-out can perhaps be equally mitigated by restricting the capabilities of a remote shell, as by a remote desktop connection with copy-paste disabled. Complete protection from scripted screen-capture copy-out is equally impossible in both cases. 

Blocking commands, including those referred to above, commonly used for data copy-out, 
is therefore a per-tier configuration question. For detail on recommendations at each tier, see section~\ref{sec:choices}.

We note, however, that basic steps to provide for more secure remote shell access should be implemented. At every tier, secure passphrases (see section 5 of the NCSC's guidance~\cite{NCSCsecurepasswords}) or public key authentication
should be enforced, with users trained in the use of key-chain managers on their access devices, locked with two-factor authentication, so that the inconvenience of repeatedly
typing a long passphrase is mitigated, reducing the risk of users choosing insecure passwords.

If implementers choose to allow ssh access at higher tiers, then specific commands, including those referred to above, commonly used for data copy-out, must be forbidden to users. With these restrictions in place, we believe that allowing secure shell access to analysis environments, from managed devices on managed networks, can provide equivalent risk of data copy-out to remote desktop access. 

\subsection{User device networks}

Our recommended network security model requires research organisations to create two dedicated research networks for
user devices, in addition to the open internet:

\begin{itemize}
    \item An Institutional network
    \item A Restricted network
\end{itemize}

An Institutional network corresponds to organisational guest network access (such as Eduroam). Access to environments can be restricted such that access is only allowed by devices which are connected to an approved Institutional network, but it is assumed the whole research community can access this network, though this access may be remote for authorised users (for example, via VPN). Restriction
by IP address is possible on the Institutional network.

Remote access to an institutional network, e.g. via VPN, may be permitted.

Access to a restricted Restricted network should be limited to a physical space of a particular security level (see section \ref{sec:physicalsecurity}). Restricted networks therefore provide a way to use restrictions on the IP ranges that can connect to higher Tier environments to enforce physical security policies.

A Restricted network may be linked between multiple institutions (such as partner research institutions), so that researchers travelling to collaborators' sites will be able to connect to Restricted networks, and thus to secure environments, while away from their home institution.

Remote access to a Restricted network (e.g. via VPN) should not be possible.

Firewall rules for the environments must enforce Restricted network IP ranges corresponding to these networks.

Of course, environments themselves should, at some tiers, be restricted from accessing anything outside an isolated network for that secure environment.

\subsection{Physical security}
\label{sec:physicalsecurity}

Some data requires a physical security layer around not just the data centre,
but the physical environment users use to connect to it.

We distinguish three levels of physical security for research spaces:

\begin{itemize}
    \item Open research spaces
    \item Medium security research spaces
    \item High security research spaces
\end{itemize}

Open research spaces include university libraries, cafes and common rooms.

Medium security research spaces control the possibility of unauthorised viewing.
Card access or other means of restricting entry to only known researchers (such as the signing in of guests on a known list) is required. Screen adaptations or desk partitions can be adopted in open plan environments if there is a high risk of ``visual eavesdropping''. Screens must be locked when the user is away from the device.

Secure research spaces control the possibility of the researcher deliberately
removing data. Devices will be locked to appropriate desks, and neither enter nor leave 
the space. Mobile devices should be removed before entering, to block the 'photographic hole',
where mobile phones are used to capture secure data from a screen. Only researchers associated with a secure project should have access to such a space.

Firewall rules for the environments must enforce Restricted network IP ranges corresponding to these 
research spaces.

\section{User lifecycle}

Having described the elements of a secure data science research environment, we now describe our recommendations for the processes which govern them.

Users who wish to have access to the environment should create credentials within the upstream infrastructure
provision service. This process will normally be managed through the creation of the user accounts within an
organisational directory, and these credentials will be proxied by the management system.

Projects are created in the management system by a Programme Manager, and an Investigator and Project Manager assigned.

Programme Managers can assign users to all projects, and can assign themselves as Project Managers. They add users to groups corresponding to specific research projects through the management framework. The Project Manager has oversight over a particular project, and can assign existing users to their project, but not invite new users to the system as a whole - only Programme Managers can do this. 

At some tiers, new members of the research team or Referees must also be approved by the Dataset Provider Representative (see~\ref{sec:choices}).

Before joining a project, Researchers, Investigators and Referees must receive training in handling data in the system and certify this in the management system.

As required by law, the Dataset Provider Representative must also certify that the organisation providing the dataset has permission from the dataset owner, and, in relation to personal data, that the lawful basis for its collection was not in itself consent,
if they are not the dataset owner, to share it with the research organisation, and this should be recorded within the management system database~\cite{GDPR}.

\section{Software Ingress}

As discussed above, the base analysis machine provided in the secure research environments should come with a side range of common data science software pre-installed. Package mirrors also allow access to a wide range of libraries, where package repositories are available for the language and mirrors are provided by the environment.

For languages for which no package mirror is provided, or for software which is not available from a package repository, an alternative method of software ingress must be provided. This includes custom researcher-written code not available via the package mirrors (e.g. code available on a researcher's personal or institutional Github repositories).

For Tier 0 and Tier 1 environments, the analysis machine has outbound access to the internet and software can be installed in the usual manner by either a normal user or an administrator as required.

For Tier 2 environments and above, the following software ingress options are available.

\subsection{Installation during virtual machine deployment}

Where requirements for additional software are known in advance of an analysis machine being deployed to a secure research environment, the additional software can be installed during deployment. In this case, software installation is performed while the virtual machine is outside of the environment with outbound internet access available, but no access to any work package data. Once the additional software has been installed, the virtual machine is ingressed to the environment.

\subsection{Installation after virtual machine deployment}

Once a machine has been deployed into an environment and had access to the data within it, it is difficult to securely and reliably move the machine out of the environment while ensuring none of the data comes with it. Therefore we recommend that analysis machines remain with the environment once they have been deployed there and are never moved out. As Environments at Tier 2 and above do not have access to the internet, any additional software required must therefore be brought into the Environment in order to be installed.

Software is ingressed in a similar manner as data, using a software ingress volume:

\begin{itemize}
    \item In external mode the Researcher is provided temporary write-only access to the software ingress volume.

    \item Once the Researcher transfers the software source or installation package to this volume, their access is revoked and the software is subject to a level of review appropriate to the environment tier.

    \item Once any required review has been passed, the software ingress volume is switched to internal mode, where it is made available to Researchers within the environment with read-only access.
\end{itemize}

For software that does not require administrative rights to install, the Researcher can then install the software or transfer the source to a version control repository within the environment as appropriate.

For software that requires administrative rights to install, the a System Manager must run the installation process.

\section{Classification Review}
\label{sec:review}

Dataset Providers feel less acutely the productivity loss of over-classification, and can attribute slow delivery to the capabilities of the research team rather than the nature of working in a secure environment. Researchers, conversely, shoulder less of the risk from under-classification. 

Therefore, we recommend that an independent Referee is consulted when classifying a work package which is likely to be Tier 2 or above.  

The Referee should classify the work package separately to the research team and Dataset Provider(s). All parties should agree on a classification in order to proceed.

This review should also take place at the point of data egress, whether for further analysis or publication (at Tier 0). At the point of egress, Referees should be given access to the Output Volume to check the data and the analysis script, and they may ask for further clarification on the script or the structure of the data. Access to the script is an essential part of classifying derived data.

For review at declassification, some implementers may wish to construct things so that the Referee can access the environment with the Output Volume and Software Volume mounted, but without the Secure Data Volume, where a project does not wish to allow the Referee access to the original data. 

The Referee needs to be familiar with the classification process outlined here, and particularly the important of considering linked data in work packages during classification – the web framework should draw the Referee’s attention to the salient literature on this topic. 

Procedures for appointing a Referee are outlined below.

\subsection{Referee college}

We believe that a peer-based review system is critical to the integrity of data classification. We therefore recommend dedicating appropriate time and resource to establishing a pool of competent Referees who can support with the classification process and procedures outlined above.

We recommend that a college of Referees is formed from Investigators, Researchers or Programme Managers who have access to the system. When Referees are needed for specific reviews on project work, they can be drawn from this college; for this reason the membership pool should be deep enough that multiple projects can request Referees. 

We believe that it would be most practical for Referees to be drawn from this pool and assigned to particular projects, as this can save time when classifying work packages containing repeat datasets with which the Referee will already be familiar. However, it would be possible to assign Referees from the pool to individual tasks within a project, and this would have the benefit of ensuring a fresh perspective for each review.

Individuals making up the Referee pool must be:

\begin{itemize}
    \item Familiar with the general principles of data handling and classification
    \item Familiar with the specific classification process and procedures
    \item Aware of the legal or financial consequences of incorrect classification
    \item Given permissions to access the project team’s data despite not being part of the research team
\end{itemize}

A Referee working on a particular task or project must be fully independent, in the sense that their interests are not aligned with those of the project team. We appreciate that a pool of Referees working with researchers in the same institution will have connections to project team members, so we assess that it is enough for the Referee to have no immediate interest in the outcomes of the project team’s work to be independent. 

As is usual for peer-review pools, this is uncompensated labour. We accept that it may be challenging to develop this as a new norm; we hope that peer review, as is common place in academia for papers and grants, will become part of the culture for sensitive data handling. 

However, Referee pools will take time to grow. We recommend that Programme Managers within the institution should be able to act as fall-back Referees if one is not available in a timely fashion.

If a research organisation wishes to use a Referee who isn’t employed there, they should establish a data processing agreement with their home organisation and, where required by the Dataset Provider, a confidentiality agreement or security clearance. We see no problem with external Referees, and while it will take more time to establish a Referee pool that extends beyond one organisation, we believe that this is a worthy aspiration.

In some cases, the Dataset Provider might appoint their own Referee to review the derived data before it is extracted from the secure environment. We also see no issue with this, though it is beyond the scope of this paper to recommend procedures for ensuring independence in the case of Dataset Provider-nominated Referees.

\section{The choices}
\label{sec:choices}

Having described the full model, processes, and lifecycles, we can now enumerate the list of choices that
can be made for each environment. These should all be separately configurable on an environment-by-environment basis. However, we recommend the following at each tier.

\begin{figure*}[htbp]
  \centering
  \includegraphics[width=16cm, height=8cm]{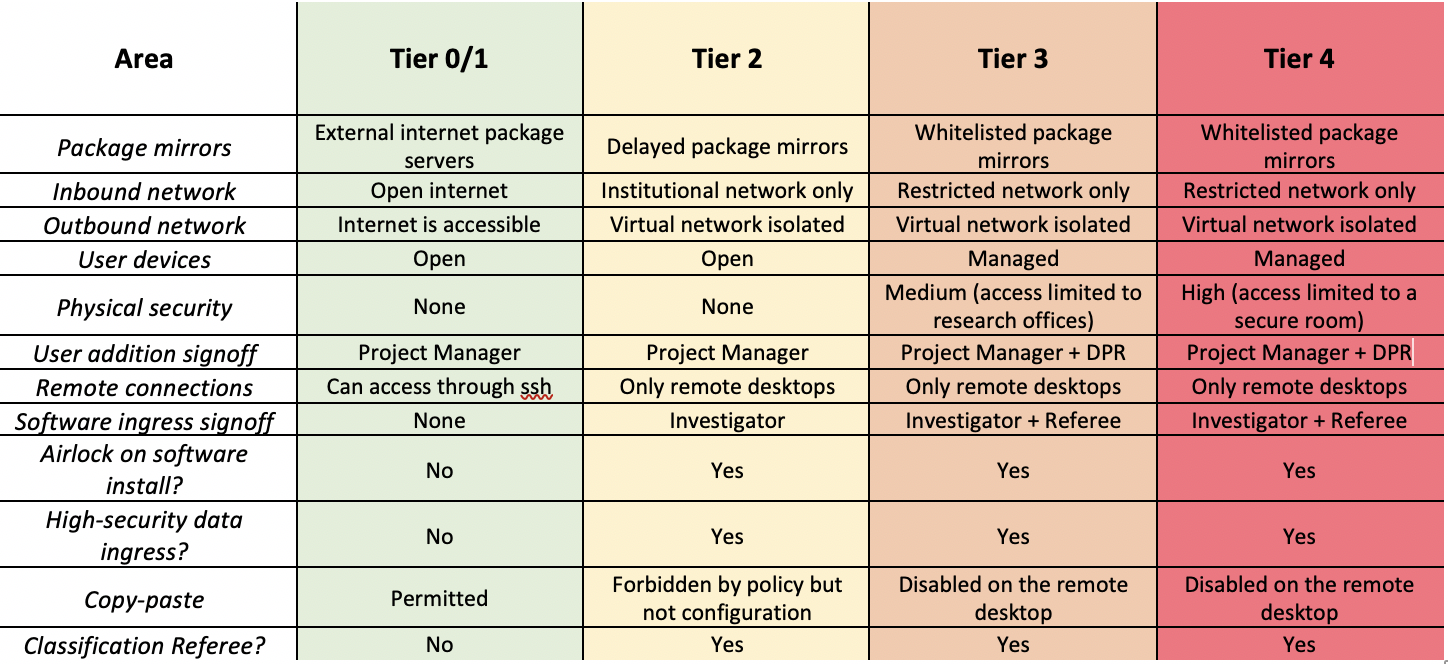}
  \caption{\label{controls} Table showing recommendations for the controls to be enacted at each tier.}
\end{figure*}

These are summarised in Figure ~\ref{controls}.

\subsection{Package mirrors}

At Tier 3 and higher, package mirrors should include only white-listed software.

At Tier 2, package mirrors should include all software, no more than 6 weeks behind the reference package server.
Critical security updates should be fast-tracked.

At Tier 1 and Tier 0, installation should be from the reference package server on the external internet.

\subsection{Inbound network}

Only the Restricted network will be able to access Tier 3 and above access nodes.

Tier 2 environment access nodes should only be accessible from an Institutional network.

Tier 1 and Tier 0 environments should be accessible from the Internet.

\subsection{Outbound network}

At Tier 1 and 0 the internet is accessible from inside the environment,
at all other tiers the virtual network inside the environment is completely isolated.

\subsection{User devices}

Open devices can access Tier 0,1 and 2.

Only managed devices can access Tiers 3 and 4.

Open devices should not be able to access the Restricted network.

Managed laptop devices should be able to leave the physical office where the Restricted network exists, but should have no access to Tier 3 or above environments while `roaming'.

\subsection{Physical security}

Tier 2 and below environments should not be subject to physical security.

Tier 3 access should be from the medium security space.

Tier 4 access must be from the high security space (see section \ref{sec:physicalsecurity}).

\subsection{User management}

The Project Manager has the authority to add new members to the research team, and the Project Manager has the authority to assign Referees.

If researchers will be working on work packages at Tier 3 and above, they must also be counter-approved by the Dataset Provider Representative, as should Referees who will be reviewing work packages of this sensitivity.

\subsection{Connection}

At Tier 1 and Tier 0, ssh access to the environment is possible without restrictions. The user should be able to set up port forwarding (ssh tunnel) and use this to access remotely-running UI clients via a native client browser.

At Tier 2 and above, only remote desktop access is enabled.

We hope it will, in future, be possible to enable secure shell access with two factor authentication and with the most common copy-out commands blocked, at Tier 2, but this remains unreliable with current technologies and we cannot recommend it without further research.

\subsection{Software ingress}

At Tier 2 and above, 
additional software or virtual machines arriving through the software
ingress process must be reviewed and signed off by the Investigator before they can be accessed inside the environment (with the exception of pre-approved virtual machines or package mirrors). At Tier 3 and higher, the Refeee should also sign this off.

For Tier 0 and Tier 1, users should be able to install software directly into the environment (in user space) from the open internet.

\subsection{Data ingress}

All data should be ingressed initially into a Tier 3 environment, unless there is any suspicion that it might require Tier 4. 

After an initial review is conducted in the Tier 3 environment, follow the protocols in section~\ref{sec:datalifecycle}.

\subsection{Copy-paste}

At Tier 1 and 0 copy-out should be permitted where a user reasonably believes their local device is secure, with the permission of the Investigator.

At Tier 2, copy-paste out of the secure research environment must be forbidden by policy, but not enforced by configuration. Users must have to confirm they understand and accept the policy on signup using the web management framework.

At Tiers 3 and 4, copy-paste is disabled on the remote desktop.

\subsection{Refereeing of classification}

Independent Referee scrutiny of data classification is required when the initial classification
by the Investigator and Dataset Provider Representative is Tier 2 or higher and for personal data which has been anonymised and is classified as Tier 1 or Tier 0.

\section{Summary of recommendations}

We recommend that organisations building environments for secure, productive data science research at scale:

\begin{itemize}
    \item Build on top of a software-defined infrastructure service
    \item Describe their entire infrastructure as a software defined infrastructure
    \item Allow such scripts to construct an independent environment for each project
    \item Define common security choices, that differ for different projects, using a configuration file interpreted by those scripts.
    \item Provide a standard set of tiers, with a clear set of questions used to determine the tier to which a
    research activity should be allocated.
    \item Classify work packages rather than datasets, taking into account the results of intended analysis and dataset combination when assessing sensitivity.
    \item Include at least one intermediate tier for pseudonymised or synthetic data, below that used for identifiable personal data, and above that used for non-personal data.
    \item Allow projects to use more than one environment, for example, where data is generated warranting use of a higher or lower tier in the course of a project.
    \item Use an independent referee college to scrutinise classification decisions
    \item Use in-environment package mirrors to allow productive import of research software libraries.
    \item Use an `airlock protocol' to allow ingress of researcher-written software
    \item Manage the associated business processes through a web-based application, backed by a database
    \item Allow this application to drive the creation and management of environments and users in the software
    defined infrastructure layer, using only forwarded credentials.
\end{itemize}

\section{Forthcoming work}

In a future paper, we hope to present a reference implementation of a cloud data analysis environment based on Microsoft Azure, compliant with these recommendations, an open source fully scripted solution allowing organisations to implement these recommendations with minimal effort.

A template Data Protection Impact Assessment to match the paper's model is in development.

\section{Declarations}

\subsection*{Author contributions}

IC began our secure research environment development activity at the Turing, and, with RC, led the creation of our first such environment.
JH proposed research towards a Turing-recommended reference secure data science research environment, coordinated the development of this design paper, obtained funding for the research, and wrote the bulk of the paper text. 
OF coordinated later contributions and wrote large sections of the text.
MOR contributed significantly to the text and led later stages of the technical work.
MOR, IC, JR, RC, HST, SCD and TD are developing the second iteration of our reference implementation.
DA, OF and CAV completed the literature review.
CL, JH, OF, ST and KW designed the data classification process.
KW and EG designed the analysis environment, and JR and MOR implemented it. 
JM helped design data ingress and egress processes.
ST reviewed the paper in relation to data protection law.
SV, FK, CL, and JM led the Turing Data Study Groups which have been the primary testing ground for our process development and reference implementation.

\subsection*{Competing interests}
All financial, personal, or professional competing interests for any of the authors that
could be construed to unduly influence the content of the article are disclosed below:

Microsoft Research provided a \$5M in-kind gift of cloud time to the Turing institute, and this gift was used in preparation of this article.

\subsection*{Grant information}
The Alan Turing Institute supported by the EPSRC grant EP/N510129/1. 
\subsection*{Acknowledgements}

We thank the following colleagues for discussions and comments on drafts of this paper: 

\begin{itemize}
    \item Mark Turner, Newcastle Digital Institute.
    \item Gerry Reilly and Susheel Varma, HDR UK
    \item Donald Scobie and Mike Jackson, Edinburgh Parallel Computing Centre
    \item Peter Macallum and Ben Thomas, UCL Research IT Services
    \item Jon Crowcroft and Mark Briers, Alan Turing Institute
\end{itemize}

We thank the wider Turing data study group ((/TU/B/000012)) participants and its organizing team for testing, feedback and being one of the focal points of this endavour.

{\small\bibliographystyle{unsrtnat}
\bibliography{sample}}

\bigskip






\end{document}